\documentclass{ws-procs9x6}
\usepackage{multirow}
\newcommand{\fxp}{FMNR$\otimes$P{\sc ythia }}
\newcommand{\fxpnospace}{FMNR$\otimes$P{\sc ythia}}

\begin{document}

\title{Beauty production using D* + $\mu$ and $\mu^+$ $\mu^-$ correlations at ZEUS}

\author{A.~E. NUNCIO QUIROZ\\
 on behalf of the ZEUS Collaboration}

\address{University of Hamburg / DESY \\
 E-mail: nuncio@mail.desy.de}  

\maketitle

\abstracts{
Events with observed $D^*$ and/or muons in the final state were used to indicate 
the presence of beauty quark production and to study the correlations 
between the particles coming from the same or different B-hadrons. 
The Next-to-Leading Order (NLO) theoretical predictions for such processes were often available only at parton 
level. A method to calculate visible level cross sections at NLO, based on an 
interface of the FMNR program to P{\sc ythia} 6.2, is applied. It uses the NLO
 description at {\it b}-quark level provided by FMNR applying a statistical reduction procedure 
(REDSTAT) that allows the link to P{\sc ythia},
 from where the description of the B-hadron decay chain is obtained.
Comparisons of the data and NLO cross sections at visible and {\it b}-quark level were found 
to be consistent and equivalent.}
\vspace*{-1cm}
\section{$D^* + \mu$ and $\mu^+ \mu^-$ analysis}
In the $D^* + \mu$ channel, an 
unlike-sign combination of both particles is observed in the same detector hemisphere 
if they come from the same {\it b}-quark,
yielding a quite pure {\it b}-sample; if they come from different {\it b}-quarks, they are present in different hemispheres. This last signature is similar to the one given by charm 
background, which produces unlike-sign $D^* \mu$ pairs in a back-to-back configuration.
The measured beauty fraction in the inclusive sample
was used to obtain the cross section in the kinematic range $p^{D^*}_{T}>1.9$ GeV, $|\eta^{D^*}|<1.5$, $ p^{\mu}_{T}>1.4$ GeV  and $-1.75<\eta^{\mu}<1.3$ as shown in Table~\ref{d*muxsections}. Details on the analyses can be found in [1,2].  
\vspace*{-0.3cm}
\section{FMNR$\otimes$P{\small YTHIA} interface}
 In addition to the theoretical NLO predictions available only at parton level, predictions at the level of visible final states are needed. The FMNR program [3] provides a framework to fragment {\it b}-quarks into B-hadrons, and simulate the decay of these hadrons by interfacing them to appropriately chosen decay spectra. However, decays to complicated final states, like $D^* \mu$ from the same B-hadron with cuts on both particles, cannot be easily implemented in this scheme. A straightforward interface of the parton level events produced by FMNR to a MC-like fragmentation and simulation chain is also not practical since those events have weights (either positive or negative) which expand over more than 8 orders of magnitude. This makes such an approach extremely inefficient because high statistics needs to be generated in order to keep statistical fluctuations low.

 These difficulties were overcome in the \fxp interface in a two-step process.
 First the reduced statistics option (REDSTAT) was applied, this is implemented as an extension to FMNR.
 The method is as follows; FMNR generates sequences of correlated parton level events (among them the ones containing divergencies so that they cancel each other). REDSTAT searches for two or more high weight events with similar kinematics, combining them into a new single event by averaging the four-momenta of the partons and summing their weights.
Events are considered to have similar kinematics when the difference in transverse momentum ($p_T$), rapidity ($y$) and azimuthal angle ($\phi$) are less than user cut values that reflect the experimental detector resolution.  
For small weight events, REDSTAT makes a random decision to keep the event depending on its weight (sampling technique).
After this procedure, the weight range is reduced to about two orders of magnitude, the number of generated events is reduced, and the NLO accuracy is preserved.

In the second step, these parton level events were interfaced to the P{\sc ythia}/J{\sc etset} [4] fragmentation and decay chain, through the ``Les Houches accord'' interface.
The initial state partons were allowed to have an intrinsic $k_T$ (typically $\sim$ 300 MeV) as implemented in P{\sc ythia}.
 Parton showering was {\em not} allowed in order to avoid double counting of higher order contributions. Since the details of the threshold treatment were found to be much more important than the choice of a particular fragmentation function, the Peterson formula with $\epsilon = 0.0035$ was used for convenience. Three approaches were considered:
\begin{itemize}
\item{Independent fragmentation in the P{\sc ythia} model. 
This was used because FMNR does not provide colour connections on a event-to-event basis, and these are not required in this model.}
\item{Fragmentation in the Lund string model.  
For this, reasonable colour connections had to be associated to each FMNR event.}
\item{Independent fragmentation scheme as provided by FMNR.
Setting the B-hadron momentum equal to the {\it b}-quark momentum before reducing it according to the Peterson formula.}
\end{itemize}
The second approach was used as default, and the other two to evaluate systematic errors.
Finally, the full decay tables and kinematics implemented in P{\sc ythia} 6.2 were used to obtain a full hadron-level event. The branching ratios were empirically corrected to correspond to those obtained from the Particle Data Group (PDG).
The Weizacker-Williams approximation with an effective $Q^2 _{max} < 25$ GeV cutoff was used to include the $\sim 15 \%$ deep inelastic scattering (DIS) contribution for a combined cross section.
\vspace*{-0.3cm}
\section{Visible beauty cross sections from $ep \rightarrow eb\bar{b}X \rightarrow D^* \mu$}
The measured visible cross section is larger than, but still compatible with the \fxp NLO prediction as shown in Table~\ref{d*muxsections}. A photoproduction subsample was selected from the inclusive sample and compared with the NLO predictions from \fxpnospace. As in the inclusive case, the prediction underestimates the measured cross section, but is compatible with the measurement within the large errors. These visible level cross sections  are also compared to the ones obtained extrapolating to {\it b}-quark level using P{\sc ythia}.  From the comparison of the ratios at visible and {\it b}-quark level, one can conclude that the extrapolation was reliable.
\vspace*{-0.2cm}
\begin{table}[ph]
\tbl{Comparison of measured and predicted cross sections. For the measured, the first error is statistical and the second systematic.}
{\footnotesize
\begin{tabular}{|crrrc|}
\hline
{} &{} &{} &{} &{}\\[-1.5ex]
{} & cross section & measured (prel.) & NLO QCD & ratio\\[1ex]
\hline
\multirow{3}{*}{Visible} &{} &{} &{} &{} \\[-1.5ex]
& Total           & $214 \pm 52^{+96} _{-84}$ pb & $72^{+20} _{-13}$ pb & $3.1^{+1.6} _{-1.7}$ \\[1ex]
& $\gamma p$ only & $159 \pm 41^{+68} _{-62}$ pb & $57^{+16} _{-10}$ pb & $2.8^{+1.5} _{-1.6}$ \\[1ex]
\hline
\multirow{2}{*}{b level} &{} &{} &{} &{}\\[-1.5ex]
& Total           & $15.1 \pm 3.92^{+3.8} _{-4.7}$ nb & $5.0^{+1.7} _{-1.1}$ nb & $3.0^{+1.3} _{-1.6}$ \\[1ex]
\hline
\end{tabular}\label{d*muxsections}}
\vspace*{-13pt}
\end{table}
\vspace*{-0.2cm}
\subsection{Comparison ZEUS - H1}
The H1 Collaboration has measured in the photoproduction regime a $D^* \mu$ cross section in a slightly different kinematic region [5], its value is shown in Table~\ref{H1xsections}, and it is compared to the ZEUS cross section, extrapolated to the H1 kinematic region using the \fxp interface. H1 and ZEUS visible cross sections are consistent.
\vspace*{-0.3cm}
\begin{table}[ph]
\tbl{Comparison of H1 and ZEUS cross section for $D^* \mu$.}
{\footnotesize
\begin{tabular}{|crcc|}
\hline
{} &{} &{} &{}\\[-1.5ex]
{} & cross section & H1 & ZEUS (prel.)\\[1ex]
\hline
\multirow{2}{*}{H1 Visible} &{} &{} &{}  \\[-1.5ex]
& $\gamma p$ only & $206 \pm 53 \pm 35$ pb & $189 \pm 48^{+80} _{-73}$ pb  \\[1ex]
\hline
\end{tabular}\label{H1xsections} }
\vspace*{-13pt}
\end{table}
\section{Visible beauty cross sections from $ep \rightarrow eb\bar{b}X \rightarrow \mu^+ \mu^-$}
\begin{figure}[ht]
  \begin{minipage}[t]{.45\textwidth}
   \centerline{\epsfxsize=5.65cm\epsfbox{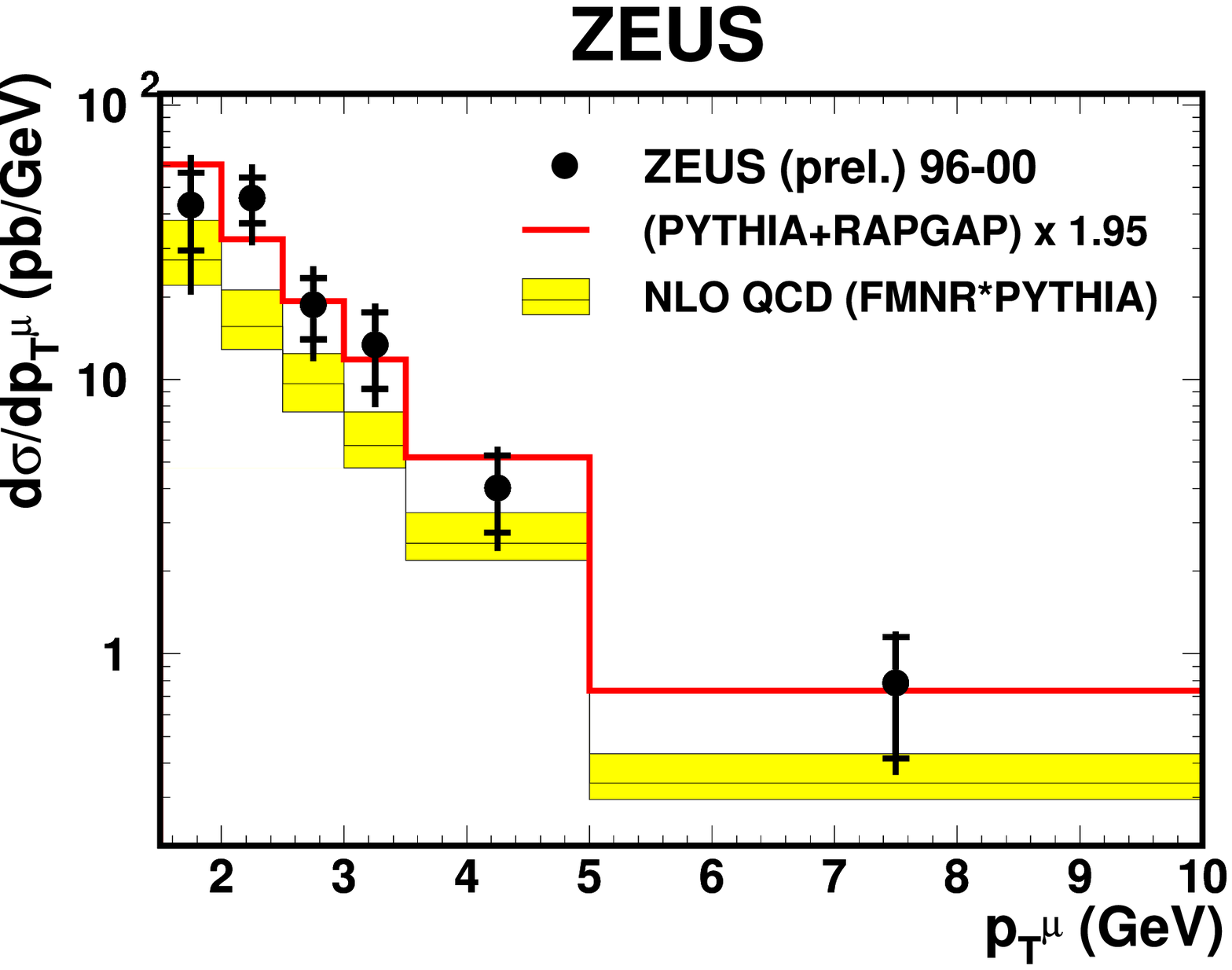}} 
  \end{minipage}
  \hfill
  \begin{minipage}[t]{.45\textwidth}
   \centerline{\epsfxsize=5.65cm\epsfbox{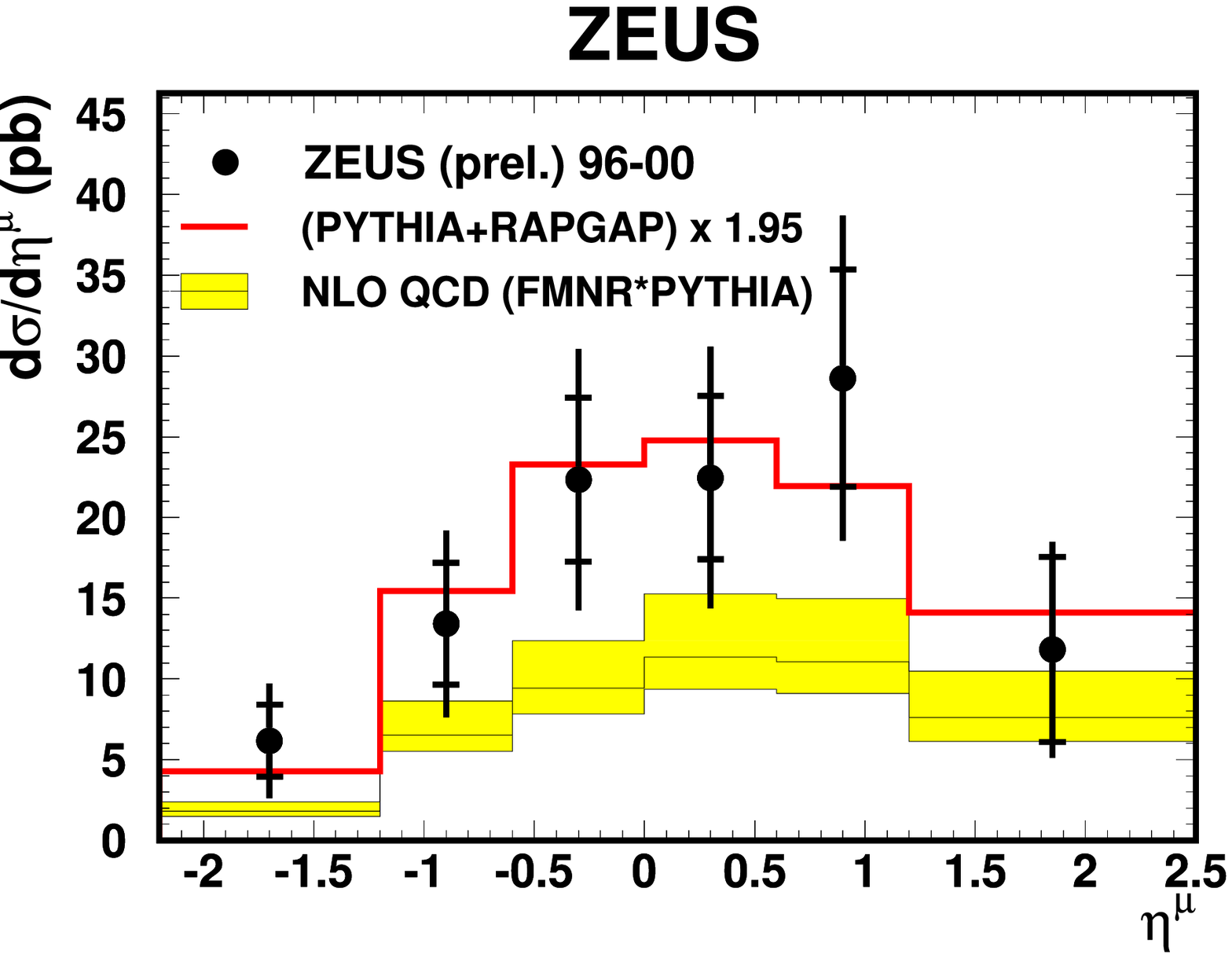}} 
  \end{minipage}
  \caption{Differential cross sections $d\sigma / dp_T$ (left) and  $d\sigma / d\eta$ (right) for muons. The data points are compared to the scaled LO prediction from P{\sc ythia} + R{\sc apgap}, and to the NLO prediction from \fxpnospace.} 
 \label{fig:pteta}
\end{figure}

For the dimuon channel, a complicated set of muon $p_T$ and $\eta$ cuts were used for maximal acceptance [2]. Table~\ref{dimuxsections} shows the measured visible cross section compared to the NLO prediction from \fxpnospace. The extrapolation to {\it b}-quark level was done using P{\sc ythia} and is compared to the NLO prediction, obtained by adding up the (standard) FMNR and HVQDIS [6] predictions, for the photoproduction and DIS regions respectively.
Here, as in the case of the $D^* \mu$ channel, the cross section comparisons at visible and {\it b}-quark level are consistent and equivalent. Differential cross sections were also obtained and shown in Figure~\ref{fig:pteta}.
\begin{table}[h]
\tbl{Comparison of measured and predicted dimuon cross sections.}
{\footnotesize
\begin{tabular}{|crrrc|}
\hline
{} &{} &{} &{} &{}\\[-1.5ex]
{} & cross section & measured (prel.) & NLO QCD & ratio\\[1ex]
\hline
\multirow{2}{*}{Visible} &{} &{} &{} &{} \\[-1.5ex]
& Total           & $63 \pm 7^{+20} _{-18}$ pb & $30^{+9} _{-6}$ pb & $2.1^{+0.8} _{-1.0}$  \\[1ex]
\hline
\multirow{2}{*}{b level} &{} &{} &{} &{}\\[-1.5ex]
& Total           & $16.1 \pm 1.8^{+5.3} _{-4.7}$ nb & $6.8^{+3.0} _{-1.7}$ nb & $2.3^{+1.0} _{-1.2}$  \\[1ex]
\hline
\end{tabular}\label{dimuxsections} }
\vspace*{-13pt}
\end{table}

\end{document}